\documentclass[conference]{IEEEtran}
\IEEEoverridecommandlockouts
\usepackage{cite}
\usepackage{amsmath,amssymb,amsfonts}
\usepackage{algorithmic}
\usepackage{graphicx}
\usepackage{textcomp}
\usepackage{xcolor}
\usepackage[hidelinks]{hyperref}
\usepackage{url}
\usepackage{pifont}
\usepackage{balance}

\def\BibTeX{{\rm B\kern-.05em{\sc i\kern-.025em b}\kern-.08em
    T\kern-.1667em\lower.7ex\hbox{E}\kern-.125emX}}
\begin{document}

\title{P4OMP: Retrieval-Augmented Prompting for OpenMP Parallelism in Serial Code
}

\author{\IEEEauthorblockN{Wali Mohammad Abdullah}
\IEEEauthorblockA{\textit{Mathematics \& Information Technology} \\
\textit{Concordia University of Edmonton}\\
Edmonton, Alberta, Canada \\
wali.abdullah@concordia.ab.ca}
\and
\IEEEauthorblockN{Azmain Kabir}
\IEEEauthorblockA{\textit{Computer Science} \\ \textit{University of Manitoba}\\
Winnipeg, Manitoba, Canada \\
kabira1@myumanitoba.ca}

}

\maketitle

\begin{abstract}
We present P4OMP, a retrieval-augmented framework for transforming serial C/C++ code into OpenMP-annotated parallel code using large language models (LLMs). To our knowledge, this is the first system to apply retrieval-based prompting for OpenMP pragma correctness without model fine-tuning or compiler instrumentation. P4OMP leverages Retrieval-Augmented Generation (RAG) with structured instructional knowledge from OpenMP tutorials to improve the reliability of prompt-driven code generation. By grounding generation in the retrieved context, P4OMP improves syntactic correctness compared to baseline prompting with GPT-3.5-Turbo. We evaluate P4OMP against a baseline—GPT-3.5-Turbo without retrieval—on a comprehensive benchmark of 108 real-world C++ programs drawn from Stack Overflow, PolyBench, and NAS benchmark suites. P4OMP achieves 100\% compilation success on all parallelizable cases, while the baseline fails to compile in 20 out of 108 cases. Six cases that rely on non-random-access iterators or thread-unsafe constructs are excluded due to fundamental OpenMP limitations. A detailed analysis demonstrates how P4OMP consistently avoids scoping errors, syntactic misuse, and invalid directive combinations that commonly affect baseline-generated code. We further demonstrate strong runtime scaling across seven compute-intensive benchmarks on an HPC cluster. P4OMP offers a robust, modular pipeline that significantly improves the reliability and applicability of LLM-generated OpenMP code.
\end{abstract}


\section{Introduction}

OpenMP is a standard API for shared memory parallelism in C, C++, and Fortran. It provides a high-level abstraction for expressing loop-level and task-level parallelism through compiler directives, enabling developers to exploit multicore architectures with minimal code restructuring. Although the syntax for OpenMP pragmas is relatively simple, their correct application requires careful reasoning about variable scoping, loop dependencies, data sharing, and reduction operations. This complexity often poses a barrier for non-expert programmers and makes OpenMP adoption error-prone in practice. Parallelizing real-world algorithms, such as triangle counting and clique covering in large networks, has historically required expert-guided manual tuning~\cite{abdullah2022efficient,abdullah2022sparse}. While these efforts are partially successful, they underscore the difficulty of transforming serial logic into efficient parallel code. P4OMP is motivated by the need for systems that can bridge this gap using retrieval-guided prompting to automate and accelerate such transformations.

Recent developments in large language models (LLMs), such as GPT-4, have demonstrated their ability to generate source code, including parallel programming constructs. However, when used in isolation, LLMs often produce invalid or unsafe OpenMP annotations. Without access to domain-specific knowledge, these models tend to hallucinate syntax, omit required clauses, or misapply parallel directives, leading to code that fails to compile or introduces data races. Our evaluation confirms this issue, with the baseline (GPT-3.5-Turbo without retrieval) failing to compile in 20 out of 108 benchmark cases, while P4OMP compiled successfully in all parallelizable scenarios.

To address these limitations, we present P4OMP, a Retrieval-Augmented Generation (RAG) framework that incorporates trusted OpenMP instructional content into the code generation process. P4OMP embeds tutorial-based knowledge into a vector database, retrieves context relevant to the user's input code, and constructs augmented prompts that guide the LLM to produce syntactically valid and semantically correct OpenMP code. This design enables the model to operate with explicit examples and rule-based context, improving generation quality across various program structures.

In contrast to hosted LLM services such as ChatGPT, which support file uploads but lack structured retrieval or reproducible batch workflows, P4OMP offers a scriptable and customizable system suitable for code-assist tools, educational deployments, and reproducible research. It is backed by a transparent and tunable corpus of OpenMP instructional examples and is fully automatable for integration into HPC code generation pipelines.

Our contributions are as follows.
\begin{itemize}
    \item We introduce a modular RAG framework for OpenMP directive synthesis that integrates domain-specific knowledge retrieval with prompt-driven LLM code generation.
    \item We evaluate P4OMP on 108 C++ benchmark cases collected from online forums, PolyBench, and NAS suites. P4OMP outperforms the baseline, achieving 100\% compilation success across all parallelizable cases.
    
    \item We show that P4OMP-generated code scales on HPC systems, with strong runtime improvements on compute-bound workloads.
    \item We provide a reproducible benchmark suite and OpenMP tutorial corpus to support community adoption and further experimentation~\cite{datarepo}.
\end{itemize}

The remainder of this paper is structured as follows. Section~\ref{related_work} surveys related work on LLM-based parallel code generation and OpenMP synthesis. Section~\ref{methodology} outlines the design of the P4OMP system. Section~\ref{evaluation} describes the evaluation pipeline, hardware setup, and benchmark cases. Section~\ref{results} presents correctness and performance results. Section~\ref{discussion} analyzes failure cases, discusses generalization potential, and contrasts P4OMP with compiler-based systems. Section~\ref{conclusion} concludes the paper and outlines future work. Section~\ref{data_availability} describes the publicly available dataset, codebase, and artifacts used in this work, which are accessible through our open-source repository.

\section{Related Work}\label{related_work}
Recent works, such as OMPGPT \cite{chen2024ompgpt}, OMPar \cite{kadosh2024ompar}, and LASSI \cite{dearing2024lassi} demonstrate various strategies to generate OpenMP pragmas using LLMs, either through supervised fine-tuning or feedback loops. PragFormer \cite{pragformer2025} and OMPify \cite{kadosh2023ompify} apply transformer-based models to classify and predict OpenMP directives.

CoTran \cite{jana2023cotran} uses reinforcement learning with symbolic feedback, while Alsofyani et al.\ \cite{alsofyani2024datarace} use LLMs to detect data races. The OpenMP educational effort by Yi et al.\ \cite{yi2024interactive} shows the growing role of LLMs in pedagogy.

Assessments like \cite{assessment2024llm} provide benchmarks across tools. Other relevant efforts explore CUDA-to-GPT transformation~\cite{gptcuda2024}, LLM-driven compiler synthesis~\cite{hong2024llm}, GPU kernel generation~\cite{ouyang2025kernelbench}, and test-guided synthesis pipelines~\cite{ma2025unitcoder}. Our architectural inspiration also draws from modular LLM integration frameworks such as Oracular Programming \cite{laurent2025oracular}. 
These approaches fall broadly into three categories: supervised fine-tuning for directive prediction, symbolic feedback-based synthesis, and LLM-based pedagogy. P4OMP differs by introducing a RAG-based prompting pipeline that operates at the source level and has been evaluated across a large benchmark suite to validate compilation correctness.

To our knowledge, P4OMP is the first RAG-based OpenMP code generation system evaluated on over 100 C++ benchmarks spanning real-world forums and standardized kernels, without requiring model retraining or compiler instrumentation.

\section{Methodology}\label{methodology}
P4OMP is designed as a modular retrieval-augmented code generation pipeline. Its key components are described below.

\subsection{OpenMP Knowledge Base}
We collect and curate OpenMP instructional materials, including structured tutorials, examples, and clause explanations from trusted sources such as OpenMP official guides and educational repositories~\cite{yi2024interactive}. This content is embedded using OpenAI’s \texttt{text-embedding-ada-002} model and stored in a FAISS-based vector index for retrieval.

\subsection{Semantic Retrieval}
Given a serial C/C++ code input, relevant sections of the tutorial corpus are retrieved using cosine similarity. These segments include canonical patterns (e.g., reduction, loop privatization, collapse clauses) that match the code's structure or intent.

\subsection{Prompt Construction}
\begin{figure}[ht]
  \centering
  \includegraphics[width=0.25\textwidth]{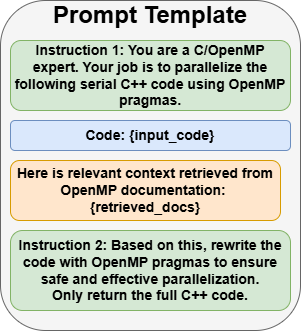}
  \caption{P4OMP prompt enriched with retrieved documentation.}
  \label{fig:p4omp_prompt}
\end{figure}

The retrieved context is integrated with the user’s serial code to construct a context-rich prompt. The prompt includes a short instruction to generate OpenMP-parallelized output, emphasizing syntactic correctness and semantic preservation (see Figure~\ref{fig:p4omp_prompt}).

\subsection{Code Generation and Validation}
The prompt is passed to GPT-3.5-Turbo via API. The generated output is then:
\begin{itemize}
    \item \textbf{Compiled} using OpenMP-enabled compilers (g++) to ensure syntactic correctness.
    \item \textbf{Validated} for semantic equivalence, such as correct loop privatization and reduction usage.
    \item \textbf{Optionally benchmarked} on local or HPC systems for runtime evaluation.
\end{itemize}

\begin{figure}[ht]
  \centering
  \includegraphics[width=0.4\textwidth]{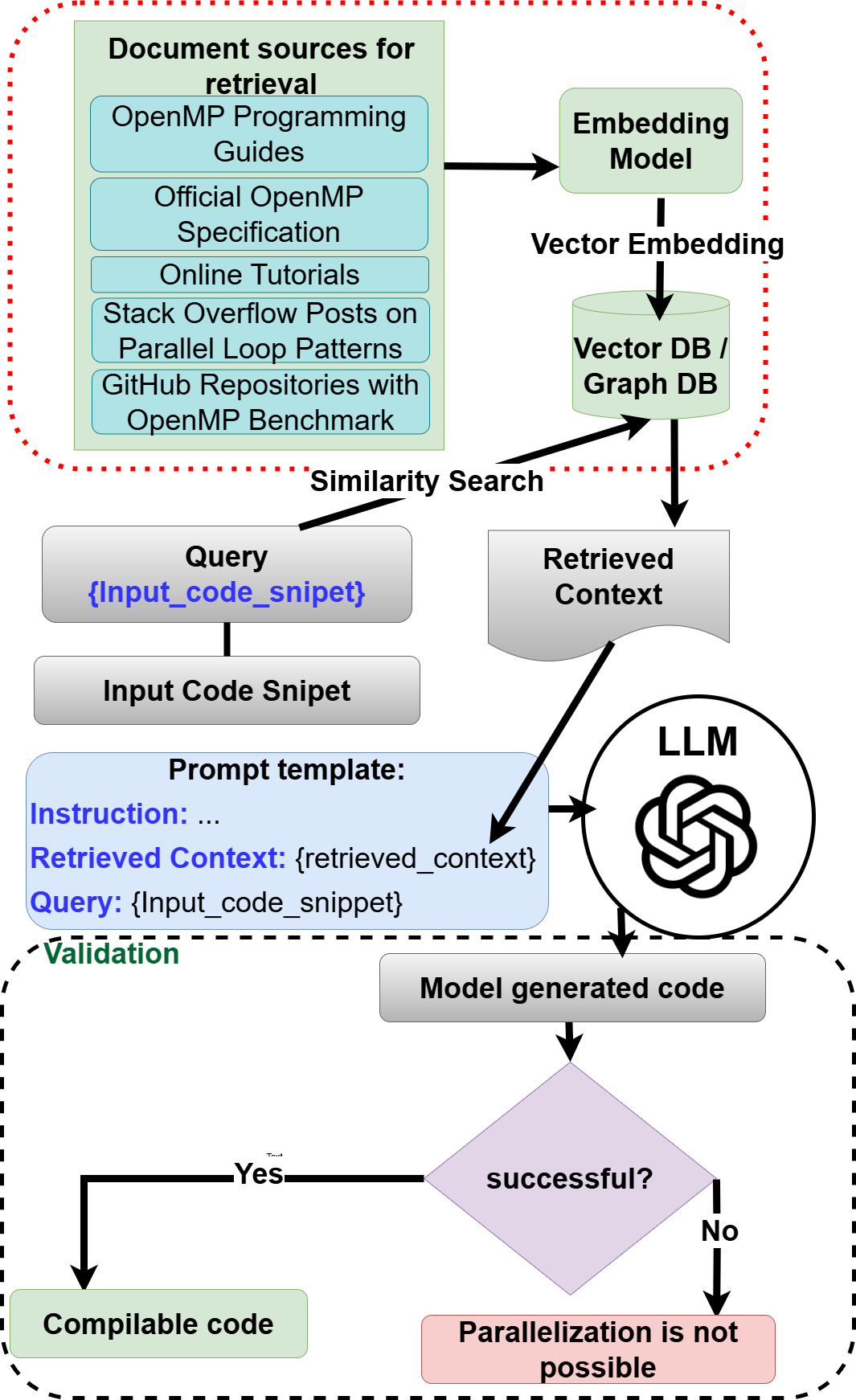}
  \caption{Overview of P4OMP architecture: semantic retrieval of tutorial content, prompt construction, and LLM-driven OpenMP code generation.}
  \label{fig:rails_methodology}
\end{figure}

Figure~\ref{fig:rails_methodology} shows the overall architecture, highlighting the retrieval-prompt-generation loop central to P4OMP. This modular flow ensures that even general-purpose LLMs can generate high-fidelity OpenMP code when guided by retrieved domain-specific context.

\section{Evaluation}\label{evaluation}
\subsection{Evaluation Setup}

We evaluated P4OMP on a dataset of 108 serial C++ programs. The first 14 cases were selected from public forums, the PolyBench suite\footnote{\url{https://www.cs.colostate.edu/~pouchet/software/polybench/}}, and NAS benchmarks\footnote{\url{https://www.nas.nasa.gov/software/npb.html}} to ensure diversity across known computational categories. We systematically collected 94 additional C++ code snippets from Stack Overflow to supplement our dataset. This was done using the Stack Exchange API with targeted keyword queries corresponding to ten algorithmic categories: dot product, matrix multiplication, quicksort, histogram, prefix sum, Jacobi 2D method, Mandelbrot set, Monte Carlo simulation, vector addition, and convolution. For each category, we retrieved up to 15 pages of accepted answers to relevant questions, sorted by relevance and vote count. From the content of the answers, we extracted code blocks and applied multiple filters to ensure quality and consistency: (i) removal of input/output operations, (ii) elimination of commented lines, (iii) enforcement of minimum structural requirements such as presence of ``\#include'' directives and valid for-loop bodies, and (iv) verification of a minimum code length. Each cleaned snippet was then tested for compilability using g++ with the C++17 standard. Only those code fragments that successfully compiled were retained in the final dataset. 

We compared the following two code generation configurations:

\begin{itemize}
\item \textbf{P4OMP}: Our Retrieval-Augmented Generation (RAG) system using \texttt{model="gpt-3.5-turbo"} with \texttt{temperature=0.2}, augmented with a structured OpenMP tutorial.
\item \textbf{Baseline}: GPT-3.5-Turbo with \texttt{temperature=0.2}, used without retrieval or any external instructional content.
\end{itemize}

For both setups, the input was a serial C++ code, and the output was an OpenMP-annotated version generated by the model. Compilation was validated for each output.

To assess runtime performance, we additionally selected 7 of these benchmarks for large-scale execution on the Graham HPC cluster under different OpenMP thread configurations (1, 2, 4, and 8 threads).

\subsection{Experiment Pipeline}
The experiment followed these steps:
\begin{enumerate}
\item \textbf{Case Selection:} We prepared 108 serial C++ files (case1.cc to case108.cc), each with loop-based parallelization potential. The first 14 (case1.cc to case14.cc) were manually curated; the remaining 94 (case15.cc to case108.cc) were automatically extracted from forum discussions and validated through preprocessing and compilation checks.
\item \textbf{Prompting:} Each case was processed by both P4OMP and the baseline configuration.
\item \textbf{Code Generation:} Outputs were generated using GPT-3.5 Turbo with \texttt{temperature=0.2}.
\item \textbf{Compilation:} We compiled the generated code using \texttt{g++ -fopenmp}.
\item \textbf{Semantic Validation:} We manually executed all 102 OpenMP-transformed programs generated by P4OMP to verify semantic correctness. For each case, outputs were compared against the corresponding serial version under both single-threaded and multi-threaded configurations to ensure behavioral consistency. This output-based validation was performed externally to P4OMP and confirmed that the transformations preserved the original program semantics.

\end{enumerate}

\subsection{Reproducibility Details}

All experiments were run on two platforms:

\begin{itemize}
\item \textbf{Local Environment:} Windows Subsystem for Linux 2 (Ubuntu 24.04), g++ 13.3, Intel Core i7-8650U CPU @ 1.90GHz, 16 GB RAM.
\item \textbf{HPC Cluster:} Experiments were conducted on the Graham supercomputing cluster, part of the Digital Research Alliance of Canada infrastructure\footnote{\url{https://www.alliancecan.ca/en}}. Each job was allocated a compute node with dual-socket Intel Xeon Gold 6130 CPUs (32 physical cores, Skylake architecture) and 64 GB RAM. The environment used g++ 12.3.1 with OpenMP enabled via the \texttt{-fopenmp} flag. Job execution was managed using SLURM with \texttt{--cpus-per-task=8} for scaling tests.


\end{itemize}

\section{Results}\label{results}
Our benchmark suite consists of 108 C++ programs. The first 14 cases span five core computational categories—linear algebra, sorting, histogramming, stencil computations, and simulations—and include programs from public forums, the PolyBench suite, and NAS Parallel Benchmarks. 
The remaining 94 cases were collected from Stack Overflow using targeted keyword-based search across ten algorithmic domains and filtered for compilation validity. 

Each program was tested using both P4OMP and baseline under identical settings (\texttt{model=gpt-3.5-turbo}, \texttt{temperature=0.2}).
Table~\ref{tab:compilation-summary} summarizes the overall compilation performance and effective coverage of both systems across the full 108-case benchmark.

\begin{table}[htbp]
\caption{Compilation Outcomes Across 108 Test Cases}
\label{tab:compilation-summary}
\centering
\begin{tabular}{|l|c|c|}
\hline
\textbf{Metric} & \textbf{Baseline} & \textbf{P4OMP} \\
\hline
Compilation Success & 82/108 (75.9\%) & 102/108 (94.4\%) \\
Failures (Fixable)  & 20/108 (18.5\%) & 0/108 (0\%) \\
Unparallelizable     & 6/108 (excluded) & 6/108 (excluded) \\
Effective Success    & 82/102 (80.4\%) & 102/102 (100\%)\\
\hline
\end{tabular}
\end{table}



Across all 108 test cases, P4OMP successfully generated compilable OpenMP code in 102 cases (94.4\%), with no compilation failures. The baseline succeeded in only 82 cases (75.9\%), failing to compile in 20 due to clause mismanagement, undeclared variables, or syntax errors. Both systems failed to parallelize 6 structurally incompatible cases due to OpenMP constraints, such as non-random-access iterators or thread-unsafe containers. When excluding these 6, P4OMP achieved 100\% compilation success on the remaining 102 cases, compared to the baseline’s 80.4\%.

\subsection{Baseline Compilation Failures Across 108 Cases}

The baseline system failed to compile 20 out of 108 test cases. These failures are not due to the limitations of OpenMP itself, but rather to incorrect or incomplete transformations generated by the LLM. The following categories summarize the root causes:

\begin{itemize}
    \item \textbf{Undeclared Variables in Clauses:} Several cases omitted necessary variable declarations in OpenMP clauses such as \texttt{private()} or \texttt{shared()}. For example, loop indices or intermediate buffers were not listed, resulting in hard compile-time errors (e.g., Cases 2, 3, 4, 11, 45, 49, 51, 61).
    
    \item \textbf{Invalid OpenMP Usage:} The baseline applied unsupported or misused OpenMP directives:
    \begin{itemize}
        \item Reductions applied to non-scalar or user-defined types (e.g., \texttt{reduction(+:C)} in Case 11).
        \item Duplicate or inconsistent use of reduction clauses (e.g., Case 95).
        \item Use of \texttt{atomic} on operations not permitted by OpenMP semantics (e.g., Case 96).
        \item Missing shared/private annotations when \texttt{default(none)} was used (e.g., Case 2).
    \end{itemize}

    \item \textbf{Syntax Errors:} Malformed pragmas, missing parentheses, or invalid loop expressions led to compiler rejections (e.g., Case 26).

    \item \textbf{Loop Collapse Misuse:} In a few cases, the baseline applied \texttt{collapse(2)} on loops that were not properly nested or indexable (e.g., Cases 71, 73).

    \item \textbf{Iterator and Range Limitations:} The generated OpenMP code attempted to parallelize over iterators that lacked random-access properties or operator overloading (e.g., Case 19 used iterators without \texttt{operator-}).

    \item \textbf{Deprecated Constructs:} In Case 79, legacy constructs such as \texttt{std::bind2nd} were misused in a parallel context, causing incompatibility with modern OpenMP-enabled compilers.
\end{itemize}


\subsection{Runtime Performance}
To assess runtime scalability, we executed seven representative P4OMP benchmarks on the Graham cluster of the Digital Research Alliance of Canada. The setup included:
\begin{itemize}
    \item 8 CPU cores (\texttt{--cpus-per-task=8}), 64 GB memory (\texttt{--mem=64G})
    \item OpenMP threads: 1, 2, 4, and 8 via \texttt{OMP\_NUM\_THREADS}
    \item Timings via \texttt{omp\_get\_wtime()} inside each benchmark
\end{itemize}

\begin{table*}[ht]
\caption{Runtime on Graham Cluster for Selected P4OMP Benchmarks with Large Inputs}
\label{tab:hpc-runtime}
\centering
\begin{tabular}{|l|c|c|c|c|c|}
\hline
\textbf{Case Name} & \textbf{Input Size} & \textbf{1 Thread (s)} & \textbf{2 Threads (s)} & \textbf{4 Threads (s)} & \textbf{8 Threads (s)} \\
\hline
Matrix Multiply (Case 2) & 2048×2048×2048 & 179.809 & 89.657 & 38.389 & 22.699 \\
Histogram (Case 4) & $10^8$ elements & 1.169 & 1.452 & 0.954 & 0.572 \\
Prefix Sum (Case 5) & $10^9$ elements & 9.579 & 5.208 & 2.773 & 1.474 \\
Jacobi 2D (Case 6) & 2048×2048 & 7.820 & 4.001 & 1.992 & 1.363 \\
GEMM PolyBench (Case 11) & 1024×1024×1024 & 32.929 & 17.664 & 8.492 & 5.228 \\
Jacobi 2D PolyBench (Case 12) & 2048×2048 & 15.978 & 8.505 & 4.295 & 2.325 \\
Conjugate Gradient NAS (Case 14) & $10^9$ elements & 32.398 & 17.556 & 10.898 & 6.935 \\
\hline
\end{tabular}
\end{table*}

\begin{figure}[ht]
\centering
\includegraphics[width=0.48\textwidth]{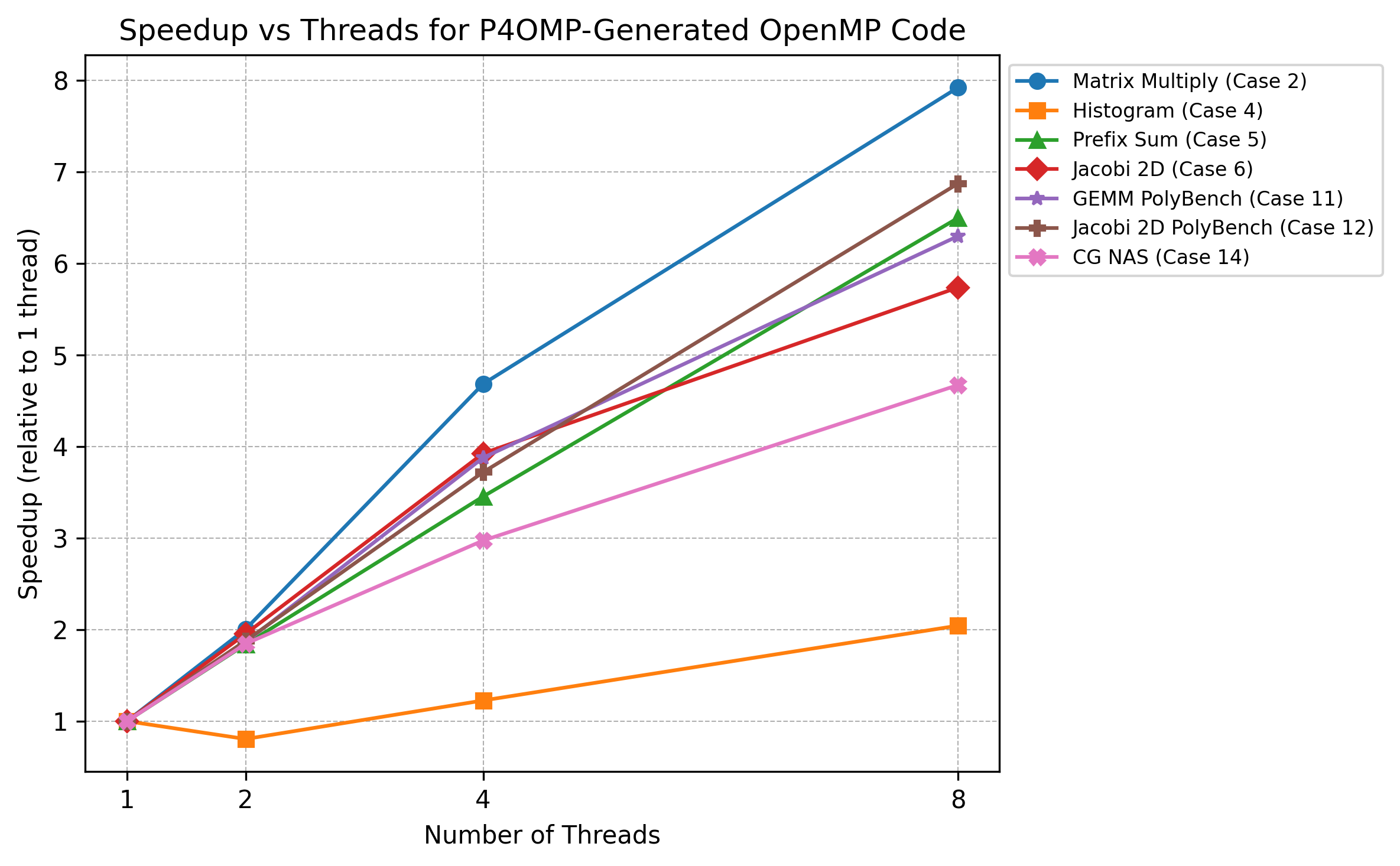}
\caption{Speedup trends for seven P4OMP benchmarks on Graham HPC cluster.}
\label{fig:hpc-speedup}
\end{figure}

Table~\ref{tab:hpc-runtime} summarizes execution times for 1, 2, 4, and 8 threads. Figure~\ref{fig:hpc-speedup} presents corresponding speedup profiles normalized to the single-thread baseline.

Compute-bound kernels, such as matrix multiplication (Case 2), Jacobi 2D (Cases 6 and 12), and conjugate gradient (Case 14), scaled efficiently, achieving speedups of up to nearly 8×. For example, Jacobi 2D (Case 6) reduced runtime from 7.82 to 1.36 seconds—a 5.7× improvement.

In contrast, memory-bound benchmarks such as histogramming (Case 4) and prefix sum (Case 5) demonstrated limited gains. Histogram showed degraded performance at 2 threads due to contention on atomic updates. This behavior is common in bandwidth-limited kernels where parallel overhead exceeds computational savings.

\textit{Memory-Bound Behavior Note:} Atomic operations and shared memory contention cause saturation in workloads like histogramming. These effects limit scalability despite correct OpenMP usage and highlight the need for workload analysis.

In summary, P4OMP-generated code compiles consistently across all evaluated benchmarks and scales well on compute-intensive workloads, enabling reproducible OpenMP parallelization without fine-tuned prompts or manual intervention.

\section{Discussion}\label{discussion}
\subsection{Tutorial Construction and Retrieval}
We manually curated an OpenMP 5.2 tutorial file integrating syntax rules, scope qualifiers, reduction strategies, and directive usage patterns.


The content was embedded using OpenAI’s \texttt{text-embedding-ada-002} model and indexed using FAISS for fast vector retrieval. Each tutorial block was separated semantically and stored to enable high-precision retrieval for diverse code contexts.

\subsection{Prompt Augmentation via Retrieval}
P4OMP augments prompts using:
\begin{enumerate}
    \item Retrieved context from OpenMP tutorials matching the user code.
    \item A static prompt template containing transformation instructions.
    \item The user's original C++ serial code.
\end{enumerate}
This method grounds the LLM in explicit, syntactically correct, and pedagogically rich OpenMP examples, preventing hallucination and scope misuse.

Although hosted LLMs support file uploads, they lack programmable workflows, retrieval customization, or integration with batch systems required in HPC contexts.

\subsection{Failure Patterns Observed in Baseline}

The baseline's failures primarily stemmed from incorrect clause usage, invalid reductions, and malformed pragmas (see Section~\ref{results})

Several cases contained malformed pragmas, missing parentheses, or incorrect loop forms—typically stemming from poor handling of iterators or complex loop structures. These patterns reflect the baseline’s lack of contextual grounding in OpenMP syntax and semantics.

By contrast, P4OMP's retrieval-augmented prompts consistently reinforced correct clause structure and directive usage, eliminating such errors across all 102 parallelizable cases.

\subsection{Complementarity with Auto-Parallelizing Compilers} 
Unlike compiler-based auto-parallelization, which operates on intermediate representations (IR) or abstract syntax trees (AST), P4OMP generates OpenMP pragmas directly at the source level, making transformations transparent and developer-friendly. This positions P4OMP as a complementary tool for education, rapid prototyping, and code assistance, rather than a replacement for production-grade compiler infrastructure.

\subsection{Lessons and Future Work}
These results underscore that retrieval is essential for high-accuracy LLM transformation. Future work includes:
\begin{itemize}
    \item \textbf{Automated Tutorial Generators:} Extracting contextually useful OpenMP examples from online documentation and repositories.
    \item \textbf{Scope Inference Assistants:} Predicting shared/private/reduction clauses from static code analysis to further aid prompt construction.
    \item \textbf{Dynamic Context Adaptation:} Currently, P4OMP retrieves tutorial chunks based on static semantic similarity. A promising future direction is to make this process dynamic—adapting the retrieved context based on intermediate feedback, code compilation outcomes, or program structure. This could enable context pruning or enrichment at runtime and improve efficiency and accuracy in large codebases.
    \item \textbf{Beyond OpenMP:} While this work focuses on OpenMP, the RAG architecture is model-agnostic and can be extended to heterogeneous computing frameworks like CUDA, SYCL, and task-based runtime.
    \item \textbf{Benchmark Expansion:} We plan to continuously expand the benchmark dataset to include real-world HPC kernels, loop nests with irregular bounds, and pointer-heavy data structures.
\end{itemize}
In future iterations, we plan to adapt retrieval context dynamically based on loop shape, variable scope patterns, or static analysis of the input code.

\subsection{Scalability and Generalization}

Our evaluation spans 108 C++ benchmark cases, including 14 curated programs and 94 additional cases collected from Stack Overflow across ten computational categories. These results demonstrate that P4OMP generalizes well across a wide range of input styles, program sizes, and algorithmic domains. Retrieval-driven prompting enables scalability. As the tutorial corpus grows to cover domain-specific and advanced OpenMP features, P4OMP can generalize to broader algorithm classes.

To validate generalization and resilience to increasing complexity, our evaluation includes GEMM and Jacobi from the PolyBench suite and Monte Carlo and Conjugate Gradient from the NAS Parallel Benchmarks.

Beyond the benchmarks considered, several graph analytic applications—such as triangle counting and clique covering—have explored partial OpenMP parallelization using intersection-based data representations~\cite{abdullah2021intersection,abdullah2022sparse,abdullah2022efficient}. These examples illustrate the difficulty of manually optimizing nontrivial algorithms. P4OMP offers a promising direction toward automating such transformations by leveraging retrieval-augmented prompting grounded in structured tutorial knowledge. While our evaluation covers 108 cases, including curated and community-sourced programs, it does not yet include large-scale industrial codebases or deeply irregular control structures. Future work will extend P4OMP's tutorial corpus and benchmarking to include complex scientific workflows, irregular graphs, and domain-specific kernels.
These directions will help further establish P4OMP as a scalable, reliable, and extensible system for source-level parallelism.

\balance

\section{Conclusion}\label{conclusion}
We presented P4OMP, a retrieval-augmented pipeline for transforming serial C++ code to OpenMP-parallelized code using large language models. P4OMP integrates structured tutorial knowledge into LLM prompts through semantic retrieval, significantly improving the correctness of generated code.

Across 108 C++ benchmark cases, P4OMP achieved 94.4\% compilation success overall, with 100\% success on all 102 parallelizable cases. The baseline configuration compiled only 82 out of 108 cases (75.9\%), failing in 20 due to clause mismanagement, undeclared variables, invalid syntax, and OpenMP misuse.

These 108 cases span a broad range of computational patterns—including handcrafted, benchmark-derived, and community-sourced programs—demonstrating P4OMP's robustness and generalization across diverse workloads.

These findings confirm that retrieval-based grounding with domain-specific knowledge significantly improves code synthesis reliability. Future work will explore automated tutorial extraction, scope inference integration, and support for additional parallel programming models.

\section{Data Availability}\label{data_availability}
All code, benchmark cases (case1.cc–case108.cc), transformation outputs, validation logs, and the retrieval-based tutorial corpus used in this work are publicly available~\cite{datarepo} to encourage future research and reproducibility. The package includes all necessary source files, OpenMP-transformed outputs, build scripts, and a README detailing setup, usage, and evaluation instructions. The full OpenMP tutorial corpus - derived from OpenMP 5.2, LLNL, Intel, GCC documentation, and academic tutorials are also included.

\section*{Acknowledgment}
This research is funded by the generous support of Concordia University of Edmonton under the Seed Grant Program.
A part of our computations was performed on the Digital Research Alliance of Canada HPC system (https://www.alliancecan.ca), and we gratefully acknowledge their support.

\bibliographystyle{IEEEtran}
\bibliography{bibliography}

\end{document}